\documentclass[reprint,amsmath,amssymb,aps,superscriptaddress]{revtex4-1}

\usepackage{ulem}

\usepackage{url}
\usepackage{comment}
\usepackage{dsfont} 
\usepackage{xcolor}
\usepackage{color}
\usepackage{graphicx} 
\usepackage{bm}
\usepackage{color}
\usepackage{braket}
\usepackage{hyperref}
\usepackage[utf8]{inputenc}
\hypersetup{colorlinks=true,linkcolor=blue,citecolor=blue}

\newcommand{\dtau}{\partial_{\tau}}

\begin{document}

\title{Microscopic dynamics and an effective Landau-Zener transition in the
quasi-adiabatic preparation of spatially ordered states of Rydberg excitations
}

\author{Andreas F. Tzortzakakis}
\affiliation{Institute of Electronic Structure and Laser, Foundation for Research \& Technology -- Hellas, 
GR-70013 Heraklion, Crete, Greece}
\affiliation{Department of Physics, National and Kapodistrian University of Athens, GR-15784 Athens, Greece}

\author{David Petrosyan}
\affiliation{Institute of Electronic Structure and Laser, Foundation for Research \& Technology -- Hellas, 
GR-70013 Heraklion, Crete, Greece}
\affiliation{A. Alikhanyan National Science Laboratory (YerPhI), 0036 Yerevan, Armenia}

\author{Michael Fleischhauer}
\affiliation{Department of Physics and Research Center OPTIMAS, University of Kaiserslautern, D-67663 Kaiserslautern, Germany}

\author{Klaus Mølmer}
\affiliation{Aarhus Institute of Advanced Studies, Aarhus University, 
H{\o}egh-Guldbergs Gade 6B, DK-8000 Aarhus C, Denmark}
\affiliation{Center for Complex Quantum Systems, Department of Physics and Astronomy, Aarhus University, 
Ny Munkegade 120, DK-8000 Aarhus C, Denmark}

\date{\today}

\begin{abstract}
We examine the adiabatic preparation of spatially-ordered Rydberg excitations of atoms in finite one-dimensional lattices 
by frequency-chirped laser pulses, as realized in a number of recent experiments simulating quantum Ising model. 
Our aims are to unravel the microscopic mechanism of the phase transition from the unexcited state of atoms
to the antiferromagnetic-like state of Rydberg excitations by traversing an extended gapless phase, and 
to estimate the preparation fidelity of the target state in a moderately sized system amenable to detailed numerical analysis. 
%We find that, in the basis of the bare atomic states, the system climbs the ladder of Rydberg excitations 
%predominantly along the strongest-amplitude paths towards the final ordered state.  
We show that, despite its complexity, the interacting many-body system can be described 
as an effective two-level system involving a pair of lowest-energy instantaneous collective eigenstates 
of the time-dependent Hamiltonian. 
The final preparation fidelity of the target state can then be well approximated by the Landau-Zener formula, 
while the nonadiabatic population leakage during the passage can be estimated using a perturbative approach 
applied to the instantaneous collective eigenstates.
\end{abstract}

\date{\today}

\maketitle

\section{Introduction}

Interacting many-body quantum systems are widely studied in many branches of physics, 
ranging from atomic nuclei and elementary particles to molecules and condensed matter.
Their phases and phase transitions are notoriously hard to simulate; 
it is still harder to simulate and study their quenched dynamics or the dynamics of quantum phase transitions.
Recently analog quantum simulators have attracted much theoretical and experimental attention \cite{QuSim2012}.
One can efficiently simulate dynamic and static properties of a many-body quantum system 
governed by a Hamiltonian $\mathcal{H}_S$ using an analog well-controlled quantum system 
with programmable interactions.
To this end, one prepares the simulator system in a well-defined but easy to initialize state 
and dynamically evolves its state vector $\ket{\psi(t)}$ subject 
to a time-dependent Hamiltonian $\mathcal{H}_S(t)$, followed by a read-out of the final state.

One of the most advanced systems for quantum simulations of spin lattice models 
is that of ultracold atoms in optical lattices or arrays of microtraps 
laser-excited to the interacting Rydberg states
\cite{Weimer2008,Pohl2010,Schachenmayer2010,Bloch2012,Bloch2015,Bernien2017,Keesling2019,Omran2019,Samajdar2020,Ebadi2021,Bluvstein2021,Semeghini2021,Ebadi2022,Kim2018,Sanchez2018,Lienhard2018,Scholl2021,Browaeys2020,Morgado2021,Petrosyan2016,Cui2017,Samajdar2020,Samajdar2021}.
The interatomic distances and interactions can be precisely controlled, while 
the long lifetimes of the Rydberg states permit the realization of coherent quantum dynamics 
of strongly-interacting many-body systems. 
In particular, such systems can realize with very high accuracy the transverse-field Ising model for $N$ spins 
described by Hamiltonian
\begin{equation}
\mathcal{H} = - \frac{1}{2} \sum_{j=1}^{N} h_j \sigma_z^{j} + \sum_{i<j}^{N} V_{ij} \sigma_z^{i}\sigma_z^{j}
- \Omega \sum_{j=1}^{N} \sigma_x^{j} , \label{eq:HamSpin}
\end{equation}
where $h_j$ is the longitudinal and $\Omega$ the transverse magnetic fields, while the spin-spin interaction
$V_{ij} \propto |x_{ij}|^{-\alpha}$ has a power-law dependence on distance $x_{ij}$ 
(e.g., $\alpha=3$ or 6 for dipole-dipole or van der Waals interactions). 
In the classical limit of $\Omega \to 0$, assuming uniform $h_j = h$ and dominant repulsive nearest-neighbor interaction $V >0$, 
the ground state of this Hamiltonian 
for large negative values of $h \ll - 4V$ is a ferromagnet with all the spins pointing down, while for smaller values 
of $-4V \lesssim h \lesssim 4V$ the ground state is antiferromagnetic. 
In the presence of a finite transverse magnetic field $\Omega \neq 0$, as the longitudinal magnetic field $h$ is slowly swept 
between the values corresponding to the ferromagnetic and antiferromagnetic ground states, the system initially prepared
in the ferromagnetic state quasi-adiabatically transitions towards the antiferromagnetic state. 
Since the initial and final states are gapped, while %in the thermodynamic limit of $N \to \infty$
the gap is closing in the vicinity of the critical point of the (second-order) quantum phase transition, 
the preparation fidelity of the target state can be estimated using the formalism 
of the quantum Kibble-Zurek mechanism \cite{Zurek2005,Dziarmaga2005,Polkovnikov2005}.
But in any finite system $N \lesssim 100$ realized in most experiments 
\cite{Bloch2015,Bernien2017,Keesling2019,Omran2019,Ebadi2021,Bluvstein2021,Semeghini2021,Kim2018,Sanchez2018,Lienhard2018,Scholl2021}, 
the gap remains finite and we cannot use the Kibble-Zurek arguments to estimate the preparation fidelity 
of the target state via an adiabatic sweep. 
The usual intuition for finite lattices with long-range interactions $V_{ij}$ has been \cite{Pohl2010,Schachenmayer2010} 
that the system attempts to adiabatically follow the ground state of the classical ($\Omega \to 0$) Hamiltonian 
climbing a ``Devil's staircase'' with different rational fractional numbers of spin excitations separated 
by energy gaps \cite{Burnell2009,Lauer2012}. 
But in order to induce the transitions between the different spin configurations, the transverse field $\Omega$ 
should have a finite amplitude.
Many experimental protocols, therefore, use a strong time-dependent (pulsed) field $\Omega(t)$ which drives 
the system along an alternative excitation path that goes through an extended gapless parameter region 
and avoids the successive populations of the lowest-energy classical spin configurations.
As discussed more quantitatively below, a ground state corresponding to a gapped antiferromagnetic phase 
can then be prepared with surprisingly high fidelity.

In this paper, to unravel the microscopic dynamics of the system, 
we consider chains of $N \lesssim 10$ laser-driven atoms with long-range (van der Waals, $V_{ij} \propto x_{ij}^{-6}$) 
interactions between the atomic Rydberg states, and examine the adiabatic preparation of 
spatially-ordered Rydberg excitations using frequency chirped laser pulses, 
as proposed theoretically in \cite{Pohl2010,Schachenmayer2010} and studied experimentally in 
\cite{Bloch2015,Bernien2017,Keesling2019,Omran2019,Ebadi2021,Bluvstein2021,Semeghini2021,Kim2018,Sanchez2018,Lienhard2018,Scholl2021}.
Our system is initially prepared in a trivial state with all the atoms in the spin-down (ground) state, and 
while the intuition behind the proposal is to evolve the system along states with a sequentially growing number 
of equidistant Rydberg excitations, we show that the system explores other, more complex states on the way to the 
target antiferromagnetic-like final state:
On the one hand, the time dependent detuning and strength of the coupling field establish adiabatic  eigenstates 
without well defined excitation number. On the other hand, the time dependent state of the system follows neither
states with increasing integer number of excitations nor the instantaneous adiabatic ground state of the full 
quantum Hamiltonian, but is prone to non-adiabatic transfer to the excited state(s) \cite{Benseny2021}.
Yet, to a good approximation, nearly all of the population loss from the adiabatic ground state 
is due to the non-adiabatic transition to the first excited state. The fidelity of state preparation can then
be estimated using the Landau-Zener formula \cite{LZ1,LZ2} for an effective two-level system in the basis 
of instantaneous adiabatic eigenstates.

The paper is organized as follows. In Sec.~\ref{sec:protocol} we review the adiabatic preparation protocol
\cite{Pohl2010,Schachenmayer2010,Bloch2015,Bernien2017,Keesling2019,Omran2019,Ebadi2021,Bluvstein2021,Semeghini2021,Kim2018,Sanchez2018,Lienhard2018,Scholl2021}. 
In Sec.~\ref{sec:prepfidelity} we present the results of numerical simulations of the
dynamics of the system and the application of the Landau-Zener theory to estimate the fidelity of preparation of 
the target final state. Our conclusions are summarized in Sec.~\ref{sec:summary}.  
In the Appendix, we briefly revisit the lowest-order adiabatic perturbation theory and illustrate it 
with a driven two-state system \cite{Benseny2021}.

\section{The adiabatic protocol}
\label{sec:protocol}

%%%%%%%%%%%%%%%%%%%%%%%%%%%%%%%%%%
\begin{figure*}[t]
  \includegraphics[width=0.9\linewidth]{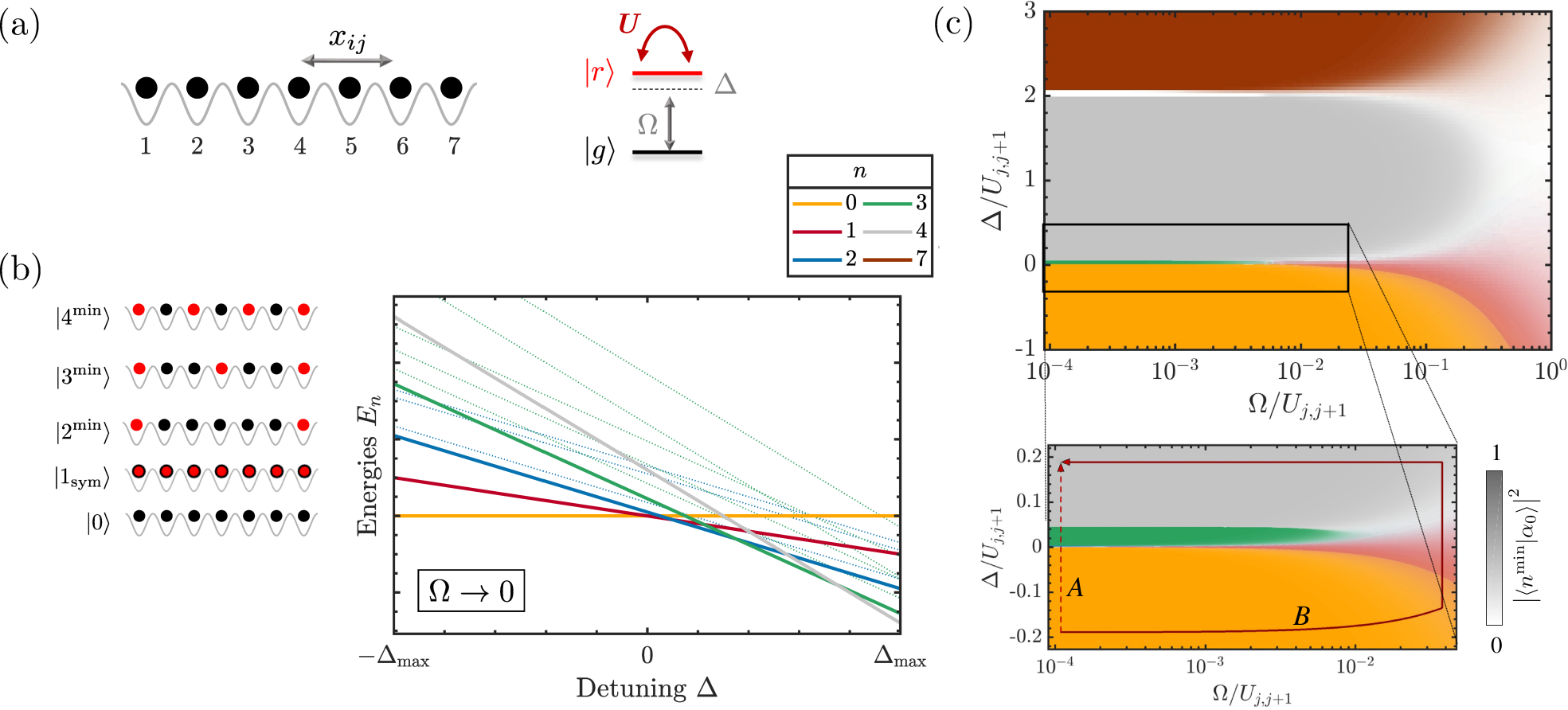}
  \caption{(a) Schematics of an array of $N=7$ atoms in a lattice. 
    The atoms interact with each other via the long-range potential $U$ when excited to the Rydberg state $\ket{r}$ 
    and are driven by a laser field on the transition $\ket{g} \to \ket{r}$ with the Rabi frequency $\Omega$ and 
    detuning $\Delta$.   
    (b)~Diagram of the energies $E_n$ of the $n=0,1,2,3,4$ excitation states versus the laser detuning $\Delta$, 
    in the limit of $\Omega\to 0$. 
    Thick solid lines correspond to the configurations $\ket{n^{\min}}$ (left) with lowest energy $E_n^{\min}$ within 
    the $n$-excitation subspace, while thin dotted lines with the same slope (and color) denote 
    the excited state energies with the same $n \geqslant 2$.
    (c)~Phase diagram of the system: 
    The upper panel shows the overlap of the ground state $\ket{\alpha_0}$
    of Hamiltonian (\ref{eq:ham}) for $N=7$ atoms with the states 
    $\ket{0}, \ket{1_{\mathrm{sym}}}, \ket{2^{\min}}, \ket{3^{\min}}, \ket{4^{\min}}$ 
    (same color code as in (b); shading intensity proportional to the overlap with each state) 
    and the fully excited state $\ket{7}$ (brown; shaded). 
    The lower panel shows the magnified diagram, with \textit{A} (dark-red dashed line) and 
    \textit{B} (dark-red solid line) denoting the classical ($\Omega\to 0$ as in Fig.~\ref{fig:IsingDiagram}(b)) 
    and quantum ($\Omega(t) \leq 0.04 U_{j,j+1}$ as in Fig.~\ref{fig:BSDyn}(a)) excitation paths.} 
  \label{fig:IsingDiagram}
\end{figure*}
%%%%%%%%%%%%%%%%%%%%%%%%%%%%%%%%%%

Consider a chain of $N$ cold atoms trapped in an optical lattice or an array of microtraps, 
as shown schematically in Fig.~\ref{fig:IsingDiagram}(a).
We treat each atom as a two-level system, with the ground state $\ket{g}$ and the excited Rydberg state $\ket{r}$ 
coupled by a spatially-uniform laser field with the time-dependent Rabi frequency $\Omega(t)$ and 
detuning $\Delta(t)=\omega(t)-\omega_{rg}$, where $\omega(t)$ is the laser frequency and 
$\omega_{rg}$ is the $\ket{g} \to \ket{r}$ transition frequency.
The atoms in Rydberg states $\ket{r}$ interact with each other via the long-range (repulsive) potential $U_{ij} = C_6/ x_{ij}^6$,
where $C_6>0$ is the van der Waals coefficient and $x_{ij}$ is the distance between atoms $i$ and $j$.
In the frame rotating with the laser frequency, the system is thus described by the Hamiltonian ($\hbar=1$)
\begin{equation}
  \mathcal{H}(t)=
  - \Delta(t) \sum_{j}^{N} \sigma_{rr}^{j}+\sum_{i<j}^{N} U_{ij} \sigma_{rr}^{i}\sigma_{rr}^{j} 
- \Omega(t) \sum_{j}^{N} (\sigma_{rg}^j+\sigma_{gr}^j)
  \label{eq:ham}
\end{equation}
with $\sigma^j_{\mu\nu}=\ket{\mu}_{j}\bra{\nu}$ being the projection ($\mu=\nu$) and transition ($\mu \neq\nu$) operators 
for atom $j$. Note that upon the substitutions $\sigma_{gg}+\sigma_{rr} = \mathds{1}$, 
$\sigma_{rr}-\sigma_{gg} = \sigma_z$, $\sigma_{gr}+\sigma_{rg} = \sigma_x$, and 
an energy shift $- \frac{1}{2}\Delta N + \frac{1}{4} \sum_{i<j}^{N}U_{ij}$,
Eq.~(\ref{eq:ham}) reduces to the Ising Hamiltonian of Eq.~(\ref{eq:HamSpin})
for spin-1/2 particles in the transverse $\Omega$ and (inhomogeneous) longitudinal
$h_j = \Delta - \frac{1}{2} \sum_{i \neq j}^{N} U_{ij}$ magnetic fields, 
interacting via the long-range potential $V_{ij} = \frac{1}{4}U_{ij}$.

The total Hilbert space $\mathbb{H}=\bigoplus_{n=0}^{N}\mathbb{H}_n$ of the system consists of $(N+1)$ subspaces $\mathbb{H}_n$ 
that span all the configurations with $n=0,1,\ldots , N$ Rydberg excitations of $N$ atoms in a lattice.
The dimension of each subspace is $\dim{\mathbb{H}_n}=  {N\choose n}$, and 
the dimension of the total Hilbert space is $\dim{\mathbb{H}}=2^N$. 
In the limit of vanishing Rabi frequency $\Omega\to 0$, the spectrum of the Hamiltonian $\mathcal{H}$ reduces to that 
of the classical Ising model in a longitudinal field $h_j$. 
Without interactions, all $n$-excitation states $\ket{n}$ would be degenerate with the energy $E_n=-n\Delta$. 
The energy of state $\ket{n=0} \equiv \ket{g g \ldots g}$ is $E_0 =0$, and the energy of all single-excitation states
$\ket{n_j=1_j} \equiv  \ket{gg \ldots r_j \ldots g}$ ($j=1,2,\ldots, N$), and their symmetric superposition
$\ket{1_{\mathrm{sym}}} = \frac{1}{\sqrt{N}} \sum_j \ket{1_j}$, is $E_1 = -\Delta$.
For $n \geq 2$, the long-range interatomic interactions $U_{ij} > 0$ partially lift the degeneracy, 
and the states with the lowest energy $E_n^{\min}$ correspond to the configurations with the largest
separation between the Rydberg excitations, which minimizes the convex interaction potential $U(x)$.
Thus, in a lattice of $N$ (odd) sites with the lattice constant $a$ and total length $l=a(N-1)$,
the lowest energy states are $\ket{2_{1,N}} = \ket{rg \ldots gr} \equiv \ket{2^{\min}}$ with 
$E_2^{\min} = -2\Delta + \frac{C_6}{(l)^6}$,
$\ket{3_{1,(N+1)/2,N}} = \ket{rg \ldots grg \ldots gr} \equiv \ket{3^{\min}}$ with 
$E_3^{\min} = -3\Delta + \frac{C_6}{(l)^6} + 2 \frac{C_6}{(l/2)^6}$, etc.
More generally \cite{Lauer2012,Petrosyan2016},
\begin{eqnarray}
E_n^{\min} &=& -n \Delta + \frac{C_6}{l^6} (n-1)^6 \sum_{k=1}^{n-1} \frac{k}{(n-k)^6}
\nonumber \\
& \simeq & -n \Delta + \frac{C_6}{a^6} \frac{(n-1)^7}{(N-1)^6},
\end{eqnarray}
where in the second line we kept the largest energy contributions due to interactions 
with the closest excited neighbors.
In Fig.~\ref{fig:IsingDiagram}(b) we show schematically the energy spectrum $\{ E_n\}$ 
of a small lattice versus detuning $\Delta$.
For $\Delta < 0$, the ground state of the system is $\ket{0}$ with $E_0= 0$.
For $\Delta > 0$, the ground state is the lowest energy $n$-excitation state $\ket{n^{\min}}$ 
with $E_n^{\min} < E_{n \pm 1}^{\min}$.
Assuming equidistant excitations, the corresponding detuning is $\Delta \simeq \frac{C_6}{2 a^6} \frac{n^7}{(N-1)^6}$

With the system initially prepared in state $\ket{0}$, as the detuning $\Delta$ is swept from some negative value
to a positive value, the energies $E_{0,1,2,\ldots, n}^{\min}$ of states $\ket{n^{\min}}$ sequentially cross
at detunings $\Delta_{n \to n+1}$ for which $E_{n}^{\min} = E_{n+1}^{\min}$.
The antiferromagnetic-like state $\ket{n^{\min}} = \ket{rgrg \ldots}$, with $n=(N+1)/2$ excitations
and energy $E_{(N+1)/2}^{\min} \simeq - \frac{N+1}{2} \Delta + \frac{N-1}{2} \frac{C_6}{ (2a)^6}$, is the lowest energy eigenstate 
of the system for the detuning $3 \frac{C_6}{(2 a)^6} \lesssim \Delta_{\mathrm{AF}} \ll 2 \frac{C_6}{a^6}$.
The protocol for the adiabatic preparation of the antiferromagnetic state of Rydberg excitations
\cite{Pohl2010,Schachenmayer2010,Bloch2015,Bernien2017,Keesling2019,Omran2019,Ebadi2021,Bluvstein2021,Semeghini2021,Kim2018,Sanchez2018,Lienhard2018,Scholl2021}
is therefore to start with all the atoms in state $\ket{gg \ldots g} = \ket{0} = \ket{\psi(t=0)}$ and some $\Delta < 0$, 
smoothly switch on $\Omega$ and slowly sweep its detuning to the final value $\Delta_{\mathrm{AF}}$, 
and smoothly switch off $\Omega$ at time $t=T$.
In the presence of a finite $\Omega \neq 0$ that couples the atomic states $\ket{g}$ and $\ket{e}$,
the crossings of levels $E_{n}^{\min}$ and $E_{n+1}^{\min}$ become avoided crossing, and the system may adiabatically 
follow the ground state $\ket{0} \to \ket{1_{\mathrm{sym}}} \to \ket{2^{\min}} \to \ket{3^{\min}} \to \ldots$ 
until it reaches the target state $\ket{\psi(T)} \to \ket{n^{\min}}$ with $n=(N+1)/2$.

This intuition presumes, on the one hand, that $\Omega$ is small enough that the picture of the classical Ising model 
with bare atomic excitation configurations remains valid (see Fig.~\ref{fig:IsingDiagram}(b) and (c) path \textit{A}), 
and, on the other hand, that $\Delta$ changes sufficiently slowly compared to the level
separation in the vicinities of the avoided crossings. 
For small $\Omega$, however, the coupling between the eigenstates at the crossings would be very weak, 
leading to a break down of the adiabatic following of the ground state, as for a quantum phase transition. 
Then the system cannot be constrained to follow the lowest energy classical configurations of the Ising model. 
In the experiments, however, the value $\Omega$ is sufficiently large (see Fig.~\ref{fig:IsingDiagram}(c) path \textit{B}) 
and the intuition based on the classical Ising model fails.
During the transfer, the system avoids populating the intermediate ground states and goes through an extended gapless phase, 
as illustrated on the phase diagram of Fig.~\ref{fig:IsingDiagram}(c) and elucidated below. 
Yet, we shall see that the end-to-end process may still succeed with only little loss of population 
from the adiabatic ground state.

\section{Preparation fidelity}
\label{sec:prepfidelity}

%%%%%%%%%%%%%%%%%%%%%%%%%%%%%%%%%%%%%%%%%%%
\begin{figure*}[t]
  \includegraphics[width=0.8\linewidth]{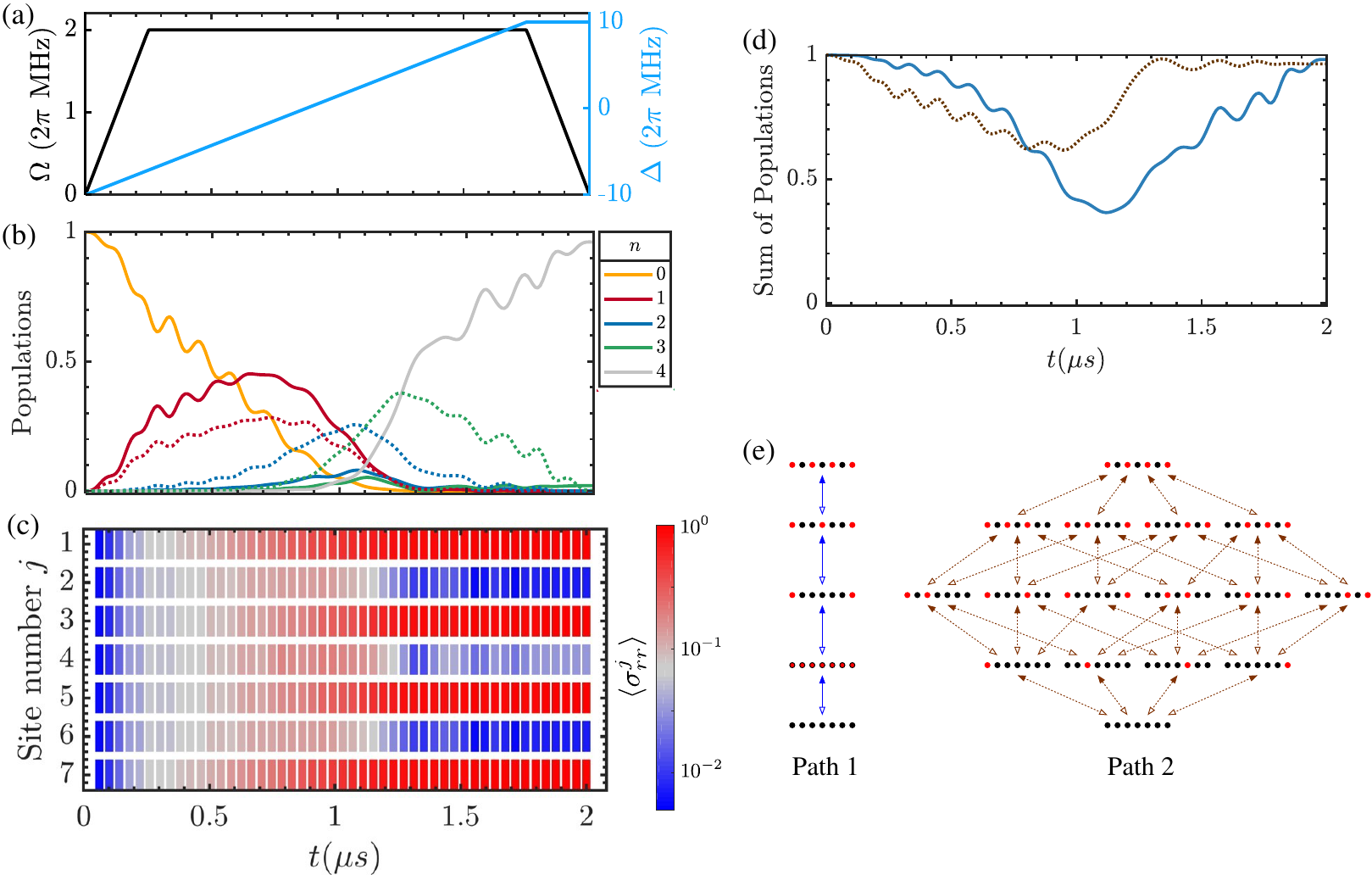}
  \caption{(a)~Time dependence of Rabi frequency $\Omega(t)$ (orange, left vertical axis) and 
    detuning $\Delta(t)$ (light blue, right vertical axis) of the laser field.
    (b)~Dynamics of populations of $n=0,1,2,3,4$ excitation states in a lattice of $N=7$ sites:
    Solid lines show the populations of states $\ket{0}$, $\ket{1_{\mathrm{sym}}}$, $\ket{2_{1,7}}$, $\ket{3_{1,4,7}}$ and 
    $\ket{4_{1,3,5,7}}$ (Path 1 in (e));
    dotted lines of the same color show the sum of populations of states $\ket{1_{1}},\ket{1_{3}},\ket{1_{5}},\ket{1_{7}}$,
    states $\ket{2_{1,3}},\ket{2_{1,5}},\ket{2_{1,7}},\ket{2_{3,5}},\ket{2_{3,7}},\ket{2_{5,7}}$, and 
    states $\ket{3_{1,3,5}},\ket{3_{1,3,7}},\ket{3_{3,5,7}}$ (Path 2 in (e)).
    (c)~Dynamics of Rydberg excitation probabilities $\braket{\sigma_{rr}^j}$ of atoms $j=1,2,\ldots,7$.
    (d)~Sum of populations of states along Path 1 in (e) (blue solid line) and of states along Path 2 in (e) (brown dotted line).
    (e)~Excitation paths of the system from state $\ket{0}$ to state $\ket{4_{1,3,5,7}}$ along the lowest energy states $\ket{n^{\min}}$ (Path 1),
    and via sequential excitation of atoms at odd lattice sites $j=1,3,5,7$ (Path 2). 
    Note that the single-excitation states $\ket{1_{1,3,5,7}}$ and the double-excited state $\ket{2_{1,7}}$ participate in both Paths 1 and 2.}
  \label{fig:BSDyn}
\end{figure*}
%%%%%%%%%%%%%%%%%%%%%%%%%%%%%%%%%%%%%%%%%

We now examine in detail the dynamics of the many-body quantum system governed by the Hamiltonian (\ref{eq:ham}).
We consider small lattices of $N\leq 9$ sites and perform exact numerical simulations \cite{Matlab}
using system parameters similar to those of recent experiments 
\cite{Bernien2017,Keesling2019,Omran2019,Ebadi2021,Ebadi2022,Bluvstein2021,Semeghini2021,Lienhard2018,Scholl2021}.
For simplicity, we consider an odd number of sites $N$ to avoid degeneracy of the target antiferromagnetic-like final state.
We assume cold $^{87}$Rb atoms with the ground state $\ket{g}\equiv \ket{5S_{1/2}, F=2, m_F=-2}$ and the exited Rydberg state
$\ket{r}\equiv\ket{70S_{1/2}, m_J=-1/2}$ coupled by a bi-chromatic laser field with the peak
two-photon Rabi frequency $\Omega_{\max}/2\pi=2 \: $MHz and 
minimum/maximum detuning $\Delta_{\min \atop \max }/2\pi=\mp 10\:$MHz.
The atoms are trapped in an array of microtraps with spacing $a \simeq 5\:\mu$m, leading to the nearest-neighbor
interaction $U_{j,j+1} \simeq 2\pi \times 53\:\textrm{MHz} \gg \Delta_{\max}, \Omega_{\max}$ and
the next-nearest-neighbor interaction $U_{j,j+2} \simeq 2\pi \times 0.8\:\textrm{MHz} < \Omega_{\max}$.

The time-dependence of the Rabi frequency and detuning of the chirped laser pulse is shown in Fig.~\ref{fig:BSDyn}(a) 
(cf. path \textit{B} in the lower panel of Fig.~\ref{fig:IsingDiagram}(c)).
The pulse duration $T \simeq 2\:\mu$s is sufficiently short to neglect the Rydberg state relaxation and 
dephasing \cite{Bernien2017,Keesling2019,Lienhard2018,Scholl2021,Petrosyan2016}, 
which permits us to consider only the unitary dynamics of the many-body system.

\subsection{Excitation configuration basis}

In Fig.~\ref{fig:BSDyn}(b) we show the dynamics of populations $P_n(t) = |\braket{n|\psi(t)}|^2$ of $n=0,1,2,3,4$ 
excitation states in a lattice of $N=7$ sites. 
The system initially in state $\ket{n=0}$ attains at the end of the pulse $t=T$ 
the target antiferromagnetic-like state $\ket{4^{\min}} \equiv \ket{4_{1,3,5,7}}$
[see Fig.~\ref{fig:BSDyn}(c)] with high fidelity $\mathcal{F} = P_4(T) \gtrsim 0.95$. 
Yet, the state vector of the system does not follow the (classical) state configurations with the lowest energies, 
which is most apparent from the low transient populations of states $\ket{2_{1,7}}$ and $\ket{3_{1,4,7}}$ 
[see Fig.~\ref{fig:BSDyn}(b)], which constitute Path 1 in Fig.~\ref{fig:BSDyn}(d,e). 
Rather, the system explores all the state configurations $\ket{2_{i,j}}$ and $\ket{3_{i,j,k}}$ 
with $i,j,k$ odd, which constitute Path 2 in Fig.~\ref{fig:BSDyn}(d,e). 
These states, despite having higher interaction energies than $\ket{2^{\min}}$ and $\ket{3^{\min}}$, 
provide a path with stronger single-photon coupling amplitudes $\propto \Omega$ to 
the final target state $\ket{4^{\min}} \equiv \ket{4_{1,3,5,7}}$ [see Fig.~\ref{fig:BSDyn}(e)].
Moreover, even though the spatially uniform laser couples the initial state $\ket{0}$ symmetrically 
to all the single-excitation states and thereby to $\ket{1_{\mathrm{sym}}}$, 
which is responsible for larger population of Path 1 than that of Path 2 at short times [see Fig.~\ref{fig:BSDyn}(d)], 
the states $\ket{1_j}$ with a single excitation on odd sites $j=1,3,5,7$ have larger populations 
than states with the excitation on even $j=2,4,6$ sites. 
Note that the transition from the lowest energy state $\ket{3^{\min}} = \ket{3_{1,4,7}}$ to the final state $\ket{4_{1,3,5,7}}$
involves a three-photon process (to annihilate the excitation on site $4$ and create two excitations on sites $3$ and $5$) 
with a correspondingly small amplitude $\propto \frac{\Omega^3}{[C_6/(l/3)^6]^2}$ \cite{Pohl2010,Petrosyan2016}. 
We then find that almost all of the reduction in the fidelity of preparation of the target state $\ket{4_{1,3,5,7}}$ 
is due to the population stuck in state $\ket{3_{1,4,7}}$. This population is also responsible for slightly larger
population of Path 1 than Path 2 at the end of the process [see Fig.~\ref{fig:BSDyn}(d)]

Thus, during the transfer, with the laser still on and its frequency being swept, the system already aims 
towards the final target state and it tends to follow the strongest-amplitude paths that lead to that state. 
While we do observe a loss of population to states with the excitations distributed in a manner that differ in several locations 
from the optimal path towards the final antiferromagnetic state, we emphasize that this loss is very limited. 
One explanation for this fact may lie in the progression of states through adiabatic passage, 
where no single intermediate state holds much occupation at any moment of time, and hence the integrated  loss from any intermediate state 
to the unwanted states is limited while almost the entire population coherently follows a progression towards the desired final state. 
We may refer to the STIRAP mechanism \cite{STIRAP} which can transfer population perfectly between two uncoupled states 
via an intermediate excited state which is minimally  populated, and hence the system suffers no loss 
despite the possible decay of that intermediate state. 

Assuming high transfer fidelity $\mathcal{F} \sim 1$ and recalling that unitary dynamics is reversible, 
we may also offer an alternative intuitive justification for the strongest-amplitude paths the system takes 
by considering a reverse process: 
If the system were initially in the antiferromagnetic state and the laser detuning was slowly swept from 
the corresponding positive $\Delta >0$ to some negative $\Delta < 0$, it would be natural to expect that 
the Rydberg excitations (initially on all the odd sites) are annihilated one at a time (and never appearing on the even sites), 
until the system reaches the zero-excitation state $\ket{0}$.
%And the system would not be likely to transition from state $\ket{4_{1,3,5,7}}$ to $\ket{3_{1,4,7}}$ via a three-photon process.
%And it would also be natural to expect that states $\ket{3_{i,j,k}}$ with odd indices $i,j,k$ to couple to states $\ket{2_{i,j}}$ with odd indices $i,j$, etc.

\subsection{Adiabatic basis}

%%%%%%%%%%%%%%%%%%%%%%%%%%%%%%%%%%%%%%%
\begin{figure}[t]
  \centerline{\includegraphics[width=0.8\linewidth]{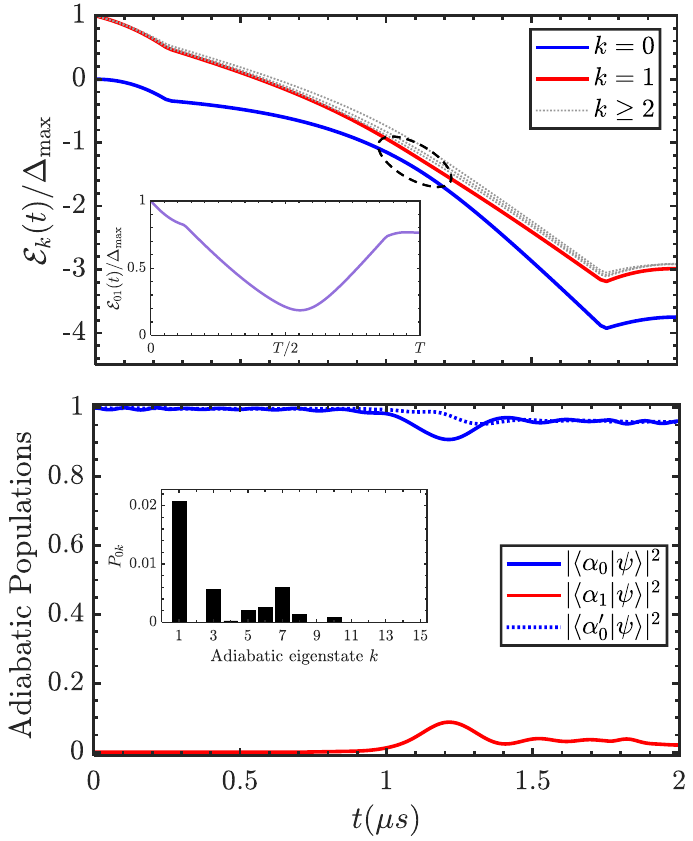}}
   \caption{Adiabatic dynamics of the system in the basis of instantaneous eigenstates $\ket{\alpha_k(t)}$ of the full Hamiltonian~(\ref{eq:ham}).
    The upper panel shows the time-dependent energies $\mathcal{E}_k(t)$ of the adiabatic states $\ket{\alpha_k(t)}$, with highlighted 
    avoided crossing between the ground $\ket{\alpha_0(t)}$ (blue solid line) and the first excited $\ket{\alpha_1(t)}$ (red solid line) state energies. 
    The inset shows the time-dependent energy gap $\mathcal{E}_{01}\equiv |\mathcal{E}_0-\mathcal{E}_1|$. 
    The lower panel shows the time-dependent populations of the ground $\ket{\alpha_0}$ (blue solid line) and first excited $\ket{\alpha_1}$ 
    (red solid line) instantaneous eigenstates. 
    Also shown is the population $|\braket{\alpha_{0}'|\psi}|^2$ of the dressed ground state $\ket{\alpha_{0}'}$ of Eq.~(\ref{eq:dress}) (blue dotted line).
    The inset shows the integrated transition probabilities $P_{0k}$ from the adiabatic ground $\ket{\alpha_0}$ to the excited states $\ket{\alpha_k}$ 
    as per Eq.~(\ref{eq:P0k}). }
  \label{fig:addyn}
\end{figure}
%%%%%%%%%%%%%%%%%%%%%%%%%%%%%%%%%%%%%%%

It is instructive to consider the dynamics of the system in the basis of instantaneous eigenstates $\ket{\alpha_k}$ 
of the full quantum Hamiltonian of Eq.~(\ref{eq:ham}),
\begin{equation}
\mathcal{H}(t)\ket{\alpha_k(t)}= \mathcal{E}_{k}(t)\ket{\alpha_k (t)},
\end{equation}
where $\mathcal{E}_{k}(t)$, with $k= 0,1, \ldots, (2^N-1)$, are the corresponding time-dependent energies.
In Fig.~\ref{fig:addyn} we illustrate the dynamics of the atomic lattice of $N=7$ sites in the adiabatic basis.
As the detuning is swept from $\Delta_{\min} <0$ to $\Delta_{\max} > 0$, the energies of the adiabatic states vary,
and the ground state $\ket{\alpha_0}$ exhibits an avoided crossing with a manifold of closely spaced excited states $\ket{\alpha_{k\geq1}}$ 
with the minimum of the energy gap $\mathcal{E}_{01}\equiv |\mathcal{E}_0-\mathcal{E}_1|$ located at a small positive value of the detuning $\Delta$. 
As expected, the system mostly resides in the adiabatic ground state, with small population of the excited states.
The largest deviation from $\ket{\alpha_0 (t)}$ occurs at the avoided crossing between the energy levels, 
after which part of the population returns from the excited states to the ground state as the levels $\mathcal{E}_{0}$ and $\mathcal{E}_{k \geq 1}$ 
separate from each other for increasing $\Delta$. This reversal of population to the adiabatic ground state is a consequence of the 
non-adiabatic time-dependent coupling between the instantaneous eigenstates, as detailed in \cite{Benseny2021} and outlined in the Appendix. 
The non-adiabatic coupling can be treated as a perturbation that dresses the adiabatic ground state by adding components of the
higher lying energy states $\ket{\alpha_{k\geq1}}$,
\begin{equation}
  \ket{\alpha_0'}  \simeq 
  \left[\ket{\alpha_0} - i \sum_{k\ne 0} \frac{ \braket{\alpha_0|(\partial_t \mathcal{H})| \alpha_k}}{(\mathcal{E}_0-\mathcal{E}_k)^2}
    \ket{\alpha_k} \right].
  \label{eq:dress}
\end{equation}
This equation indicates that the dressing $\propto (\mathcal{E}_0-\mathcal{E}_k)^{-2}$
vanishes away from the avoided crossing where the energy level separation is large. 
Hence, the excited state components of the dressed state $\ket{\alpha_0'}$ in Eq.~(\ref{eq:dress}) 
are only populated near the crossing and nearly vanish as the dressed state approaches 
the adiabatic ground eigenstate $\ket{\alpha_{0}}$ both at the initial and final times.

%%%%%%%%%%%%%%%%%%%%%%%%%%%%%%%%%%%%%%%%
\begin{figure}[t]
  \centerline{\includegraphics[width=0.8\linewidth]{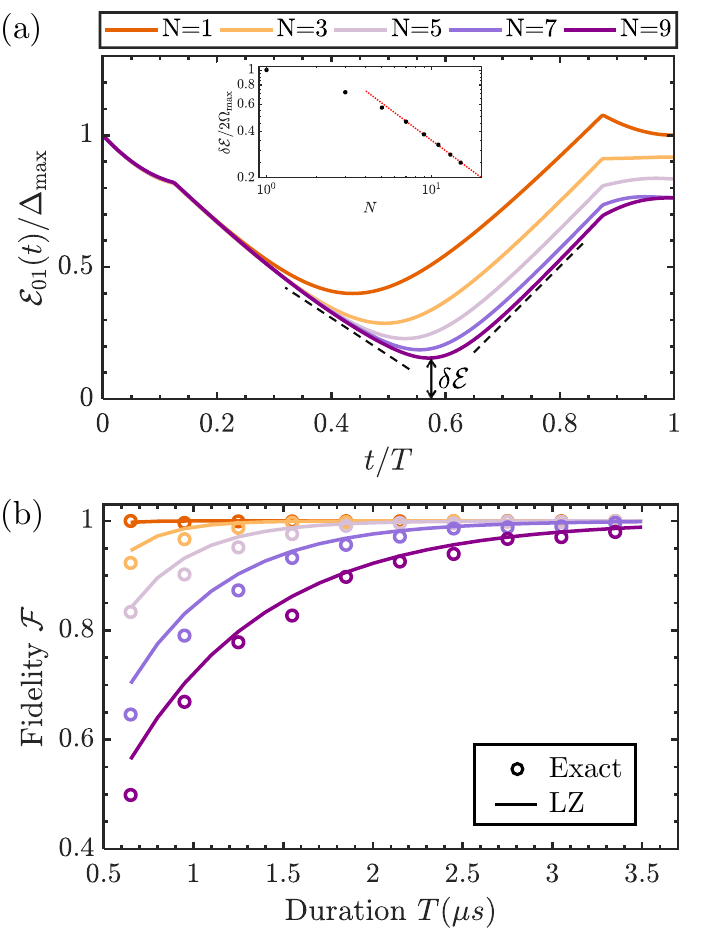}}
  \caption{(a)~Energy gap $\mathcal{E}_{01}(t)$, between the ground $\ket{\alpha_0(t)}$ and the first excited $\ket{\alpha_1(t)}$ 
    instantaneous eigenstates, as a function of the normalized time $t/T$ for different number of atoms $N$ in a lattice 
    with a fixed lattice constant $a$. 
    For each $N$, the $|\dot{\mathcal{E}}_{01}|$ in Eq.~(\ref{eq:fidel}) is calculated as the mean of the absolute values of 
    the derivatives $\dot{\mathcal{E}}_{01}|_{t_{\pm}}$ taken at $t_{\pm} = t_{\min} \pm 0.15 T$ on both sides of the corresponding 
    avoided crossing $\delta\mathcal{E}$ at $t_{\min}$, as illustrated with dashed lines for $N=9$. 
    The inset shows the minimal gap $\delta\mathcal{E}$ as obtained from exact diagonalization 
    of the Hamiltonian~(\ref{eq:ham}) for odd $N=1,3,\ldots,15$ (black circles), and 
    an asymptotic fit $\delta\mathcal{E}/(2 \Omega_{\max}) \simeq 2.2182/N^{0.8014}$ (red dotted line). 
    (b)~Preparation fidelity $\mathcal{F}=|\braket{n^{\min}| \psi(T)}|^2$ of the antiferromagnetic-like state, $n=(N+1)/2$, 
    as a function of pulse duration $T$, as obtained from the exact numerical simulations (open circles) and 
    the Landau-Zener approximation of Eq.~(\ref{eq:fidel}) (solid lines of the same color).}
  \label{fig:fid}
\end{figure}
%%%%%%%%%%%%%%%%%%%%%%%%%%%%%%%%%%%%%%%%%%

Figure~\ref{fig:addyn} also reveals that during the process most of the population lost 
from the adiabatic ground state $\ket{\alpha_0}$ goes to the first excited state $\ket{\alpha_1}$,
which, at $t=T$ ($\Omega \to 0$), coincides with the lowest-energy triple excitation state, $|\braket{3_{1,4,7}|\alpha_1(T)}|^2 =1$,
and is responsible for the reduction of the preparation fidelity of the target state.
At the same time, the contribution of the higher excited adiabatic states to the dynamics is even smaller, as attested by 
the (integrated) transition probabilities from the ground state $\ket{\alpha_0}$ to the excites states $\ket{\alpha_k}$,
\begin{equation}
P_{0k} = \left| \int_0^T \! e^{-i \int_0^t dt' \mathcal{E}_{01}(t') }  \braket{\alpha_0|\partial_t | \alpha_k} dt \right|^2  \label{eq:P0k}
\end{equation}
that follows from Eq.~(\ref{eq:main}) with $|a_0(t)| \simeq 1 \, \forall \, t \in [0,T]$.
This suggests that our strongly-interacting many-body system can be treated as an effective two-level system. 
We can then estimate the final preparation fidelity $\mathcal{F}$ of the
target antiferromagnetic-like state $\ket{\alpha_0 (T)} = \ket{n^{\min}}$ ($n=(N+1)/2$) using the Landau-Zener formula
\begin{equation}
  \mathcal{F} \simeq 1- \exp\left[- 2 \pi \frac{(\delta \mathcal{E}/2)^2}{|\dot{\mathcal{E}}_{01}|}\right],
  \label{eq:fidel}
\end{equation}
where $\delta \mathcal{E}\equiv \min[\mathcal{E}_{01}]$ is the smallest energy gap between the ground and first excited states at the avoided crossing, 
and $\dot{\mathcal{E}}_{01}$ is the time derivative of $\mathcal{E}_{01}$ taken in the interval with constant $\Omega = \Omega_{\max}$ 
but away from the avoided crossing (see Fig.~\ref{fig:fid}(a)). 
In the inset of Fig.~\ref{fig:fid}(a) we show the minimum energy gap obtained from exact diagonalization 
of the Hamiltonian~(\ref{eq:ham}) for different sizes $N \leq 15$ of the system \cite{Matlab}. 
Since during the transfer the system passes through an extended phase, which is gapless in the thermodynamic limit, 
the minimum gap decreases with increasing system size following a power law for larger $N$,
\begin{equation}
    \delta \mathcal{E} \sim N^{-\nu},
\end{equation}
where we estimate via exact diagonalization of the Hamiltonian~(\ref{eq:ham}) that $\nu\approx 0.8$, i.e., the minimum gap closes slower than $1/N$.
Thus the target state preparation can be adiabatic only in a finite system.  

In Fig.~\ref{fig:fid}(b) we show the final preparation fidelities $\mathcal{F}$, as obtained from Eq.~(\ref{eq:fidel}) and 
from exact numerical simulations, for lattices of $N=1,3,5,7,9$ sites.
We observe that, even though the transfer process has a finite duration and $\Delta_{\min \atop \max }$ are bounded, while the Landau-Zener
theory is strictly speaking applicable in the asymptotic limit of $\Delta_{\min \atop \max } \to \mp \infty$, it still gives reasonably
good estimates for the preparation fidelity of the target state in a finite system with non-vanishing energy gap $\delta \mathcal{E}$. 
Hence, achieving high transfer fidelity $\mathcal{F} \sim 1$ for larger systems $N \gg 1$ requires sufficiently long transfer times 
$T \gtrsim (\delta \mathcal{E})^{-1}$. Since the minimal energy gap closes only slowly with increasing system size, the spatially ordered 
phase of Rydberg excitations of atoms can be prepared with sizable fidelity even for moderately large atomic lattices $N \lesssim 10^2$ 
\cite{Bernien2017,Keesling2019,Lienhard2018}.
Thus extrapolating the minimal gap for larger $N$ as $\delta \mathcal{E}/2 \simeq 2.2 \, \Omega_{\max}/N^{0.8}$ while 
$|\dot{\mathcal{E}}_{01}| \simeq (\Delta_{\max} -\Delta_{\min})/T$ (see Fig.~\ref{fig:fid}(a)), 
with $\Omega_{\max}/2\pi =2\:$MHz, $\Delta_{\max \atop \min }/2\pi=\pm 10\:$MHz and $T \simeq 2\:\mu$s, 
we obtain $\mathcal{F} \simeq 0.05$ for $N=100$.

We note that an effective description of the many body system via a two-level Landau-Zener model 
can facilitate numerical calculations for much larger systems, for which finding the instantaneous ground 
and first excited states via the (target) density-matrix renormalization group calculations can be done 
with relative ease and high precision \cite{DMRG-RMP2005,Schneider2008,Bohrdt2018}, while calculations 
for successively higher excited states are increasingly difficult and less precise due to the onset of degeneracy.

Clearly, most of the non-adiabatic transition away from the ground state of the system occurs near the minimal gap 
in the vicinity of small positive detunings $\Delta \gtrsim 0$. Therefore in the experiments where the goal is a high fidelity 
of state preparation, the laser detuning $\Delta$ is swept not linearly in time but first fast, then slower around $\Delta \simeq 0$, 
and then faster again (``cubic sweep'') \cite{Bernien2017,Ebadi2021,Ebadi2022,Bluvstein2021,Semeghini2021}. 
But when the goal is to study the dynamics of quantum phase transition \cite{Bloch2015,Keesling2019,Ebadi2021,Lienhard2018,Scholl2021}, 
the sweep of $\Delta$ is linear in time. 
In general, however, optimal state preparation for various lattice sizes and/or configurations requires applications of optimal 
control methods \cite{Omran2019,Cui2017} that yield complicated time-dependent shapes of the amplitude $\Omega$ and frequency $\Delta$
of the preparation laser.

\section{Conclusions}
\label{sec:summary}

To summarize, we have presented a detailed analysis of the adiabatic preparation protocol 
of spatially ordered Rydberg excitations of atoms in finite lattices simulating the driven many-body quantum Ising model.
Our study detailed the microscopic dynamics of the preparation process both in the excitation configuration basis and 
in the time-dependent adiabatic basis.
In the excitation configuration basis, we found that the system climbs the ladder of Rydberg excitations
predominantly along the strongest-amplitude paths towards the final ordered state, whereas the usual intuition 
that the system attempts to adiabatically follow the ground state of the classical Ising Hamiltonian fails.
Since the dynamics of the system starts and ends in a gapped many-body state, but goes through an extended gapless phase, 
it cannot be described by the Kibble-Zurek mechanism. 
We showed that the transfer can instead be well described as a Landau-Zener avoided crossing of the two lowest instantaneous 
many-body eigenstates with a characteristic finite-size gap that scales with system size as $N^{-\nu}$ with $\nu\approx 0.8$. 
The transfer fidelity can then be accurately estimated from the Landau-Zener formula and 
can reach sizable values even for moderately sized systems with up to $N \sim 10^2$ atoms.
Thus, our analysis of a small and tractable system revealed the reasons 
behind a surprisingly high fidelity of preparation of the antiferromagnetic state of Rydberg excitations, 
as realized in a number of recent large-scale experiments that have already achieved quantum supremacy 
and cannot be exactly simulated on classical computers. 
While we focused on the preparation of anti ferromagnetic-like ($\mathds{Z}_2$ ordered) state  with the nearest-neighbor blockade of Rydberg excitations,
our analysis would also apply to atomic systems with stronger and/or longer-range (e.g. dipolar $\alpha=3$) interactions where a Rydberg excitation
of an atom suppressed the excitation of $m>1$ neighbors for the same final effective magnetic field $\Delta_{\max}$,
resulting in preparation of $\mathds{Z}_{m+1}$ ordered crystalline states \cite{Bernien2017}.

Our conjecture that the interacting many body quantum system admits an effective two-state description 
was deduced from numerical studies of a relatively small but non-trivial system. 
We note that, consistently with our observation, the recent study \cite{Ebadi2022} of the maximum independent set problem 
with Rydberg atom arrays has also found that the preparation fidelity of the target state -- corresponding 
to the optimal solution of the problem -- using the adiabatic protocol depends on the minimal energy gap 
between the adiabatic ground and first excited states via the Landau-Zener formula.
More generally, during the evolution the adiabatic eigenstates of the system are perturbed 
by diabatic corrections which are second (or higher) order in their energy separation.
This perturbation thus plays an important role near the (avoided) level crossings
or when the system is in the vicinity of the quantum phase transition where the energy gap is closing,
and it rapidly vanishes with increasing level separation. These arguments apply both
to simple two- or few-level systems \cite{Benseny2021} and to complicated many-body systems
with a dense energy spectrum as studied here.  

\acknowledgments
This work was supported by the EU QuantERA Project PACE-IN, GSRT Grant No. T11EPA4-00015 (A.F.T. and D.P.),
the Alexander von Humboldt Foundation in the framework of the Research Group Linkage Programme (D.P. and M.F.),
the Deutsche Forschungsgemeinschaft through SPP 1929 GiRyd (D.P. and M.F.) and SFB TR 185, project number 277625399 (M.F.), 
the Carlsberg Foundation through the Semper Ardens Research Project QCooL (K.M.), and 
the Danish National Research Foundation Centre of Excellence for Complex Quantum Systems, Grant agreement No. DNRF156 (K.M.).

%%%%%%%%%%APPENDIX%%%%%%%%%%%%%%%%%%%%%%%

\appendix

\section{Adiabatic passage and transition amplitudes} 
\label{App:eigdredd}

The adiabatic following suggests that one can attain with high fidelity a desired eigenstate 
of a complicated Hamiltonian by initially preparing the system in the corresponding eigenstate of 
a simpler Hamiltonian and then adiabatically evolving this Hamiltonian into the desired complicated one. 
In the limit of an infinitely slow evolution, the adiabatic theorem \cite{adth} ensures that the 
target state is reached with a fidelity $\mathcal{F}=1$, while for a process of finite duration 
the non-adiabatic transition probabilities associated with population losses are exponentially small 
in the duration of the process \cite{adprob,berry,berry1}. 

Here, we review a simple, lowest-order perturbative approach \cite{Benseny2021} to calculate
the non-adiabatic transition probabilities.  

Consider a time-dependent Hamiltonian $\mathcal{H}(t)$ changing slowly and continuously 
within a time interval $t\in[0,T]$, with $T$ being the duration of the process.
We assume that $\mathcal{H}(t)$ has a discrete spectrum for all times $t$.
We define the dimensionless time $\tau=t/T=v t \in [0,1]$, where $v\equiv 1/T$ 
determines the speed of the adiabatic passage \cite{Messiah}.
The state $\ket{\psi(\tau)}$ of a quantum system governed by Hamiltonian $\mathcal{H}(\tau)$
evolves in time according to the Schrödinger equation ($\hbar=1$)
\begin{equation}
iv \dtau \ket{\psi(\tau)}= \mathcal{H}(\tau) \ket{\psi(\tau)} . 
\label{eq:schr}
\end{equation}
At any instant of time, we can expand the state vector 
\begin{equation}
  \ket{\psi(\tau)}=\sum_m a_m(\tau) \ket{\alpha_m(\tau)}  
  \label{eq:exp}
\end{equation}
in the time-dependent (adiabatic) basis of the instantaneous eigenstates of the Hamiltonian
\begin{equation}
  \mathcal{H}(\tau) \ket{\alpha_m(\tau)}= \mathcal{E}_m (\tau) \ket{\alpha_m(\tau)} . 
  \label{eq:adiab}
\end{equation}
Substituting $\ket{\psi(\tau)}$ of Eq.~(\ref{eq:exp}) into Eq.~(\ref{eq:schr}), we obtain
\begin{equation}
  iv \dtau a_m = a_m  \mathcal{E}_m - iv \sum_n a_n \braket{\alpha_m |\dtau | \alpha_n} ,
  \label{eq:main}
\end{equation}
or, 
\begin{equation}
  iv \partial_{\tau} \ket{\psi(\tau)} = [\mathcal{H}(\tau) + V(\tau)] \ket{\psi(\tau)}, \label{eq:rot}
\end{equation}
with $\mathcal{H} = \sum_m  \mathcal{E}_m \ket{\alpha_m} \bra{\alpha_m}$ and 
$V =-iv \sum_{m,n} \braket{\alpha_m |\dtau | \alpha_n} \ket{\alpha_m} \bra{\alpha_n}$.
This equation shows that, in the adiabatic basis $\{\ket{\alpha_k }\}$, the time evolution of the system 
is also governed by a Schrödinger equation but with a new Hamiltonian $\mathcal{H}'(\tau)\equiv \mathcal{H}(\tau)+V(\tau)$. 
In this picture, the adiabatic basis $\{\ket{\alpha_k }\}$ is formally fixed, but its time-dependence is 
included in the diabatic correction $V(\tau)$ \cite{adprob,berry,berry1,Fleischhauer1999}.

Suppose that the system is initially in an eigenstate $\ket{\alpha_n}$ of $\mathcal{H}(\tau)$, such that
$\ket{\psi(0)} = \ket{\alpha_n}$. If the evolution were perfectly adiabatic, $V=0$, 
we would have $|\braket{\alpha_n |\psi(\tau)}| = 1$ at all times $\tau \in [0,1]$. 
Our objective is to determine the deviation from the adiabatic eigenstate using a perturbative approach. 
The key idea \cite{Benseny2021} is the following: Since Eq.~(\ref{eq:rot}) is also a Schrödinger equation 
with a time-dependent Hamiltonian, the system will (approximately) follow the instantaneous 
eigenstate $\ket{\alpha_n'}$ of $\mathcal{H}'$
\begin{equation}
\mathcal{H}'(\tau)\ket{\alpha_n'(\tau)} =  \mathcal{E}'_n (\tau)\ket{\alpha_n'(\tau)} , \label{eq:n1}
\end{equation}
which is related to $\ket{\alpha_n}$ via $\lim\limits_{v\to 0}\ket{\alpha_n'} = \ket{\alpha_n}$.
Since we consider adiabatic passage, we assume that the rate of change $v$ of the system parameters 
is smaller than the transition frequencies $ \mathcal{E}_n- \mathcal{E}_m$, and thus $\epsilon\equiv v/\min| \mathcal{E}_n- \mathcal{E}_m| \ll 1$.
We differentiate Eq.~(\ref{eq:adiab}) and substitute the expression for $\dtau \ket{\alpha_m}$ 
into Eq.~(\ref{eq:main}), obtaining 
\begin{equation}
iv \braket{\alpha_n|\dtau | \alpha_m} = i\frac{v}{\omega_{nm}}\braket{\alpha_n|(\dtau \mathcal{H}) | \alpha_m},
\; n \neq m ,
\end{equation}
which demonstrates that $\|V\|/\|\mathcal{H}\|=\mathcal{O}(\epsilon)$.
Hence, the difference between $\ket{\alpha_n'}$ and $\ket{\alpha_n}$ is indeed of $\mathcal{O}(\epsilon)$.
This of course does not mean that $\ket{\psi}$ is closely following $\ket{\alpha_n'}$ or $\ket{\alpha_n}$; 
this remains to be verified numerically. We can, however, obtain an approximate analytic expression for 
$\ket{\alpha_n'}$ by treating $V(\tau)$ as a perturbation to $\mathcal{H}$. Thus, to first order in 
$V/\mathcal{H}$ we obtain the perturbed, or ``dressed'', state  
\begin{eqnarray}
  \ket{\alpha_n'} & \simeq \frac{1}{\sqrt{M}} 
  \left[\ket{\alpha_n} -iv \sum_{m \ne n} \frac{\braket{\alpha_n |\dtau | \alpha_m}}{\mathcal{E}_n-\mathcal{E}_m} \ket{\alpha_m} \right] \\
  & = \frac{1}{\sqrt{M}} 
  \left[\ket{\alpha_n} - iv \sum_{m\ne n} \frac{ \braket{\alpha_n|(\dtau \mathcal{H})| \alpha_m}}{(\mathcal{E}_n-\mathcal{E}_m)^2} 
\ket{\alpha_m} \right] , \quad
\label{eq:corr}
\end{eqnarray}
where $M$ is an appropriate normalization constant. 
The above expression provides an intuitive picture for the non-adiabatic population losses during the passage, 
clearly indicating their strong [$\propto (\mathcal{E}_n-\mathcal{E}_m)^{-2} $] dependence on the energy level separation. 
Moreover, Eq.~(\ref{eq:corr}) provides a simple method for the calculation of the non-adiabatic correction to the adiabatic 
eigenstate during the transfer, e.g., when $\mathcal{H}'$ is a large, and difficult to diagonalize, matrix.

We could of course calculate the second and higher-order corrections to the adiabatic basis,
or diagonalize the Hamiltonian $\mathcal{H}'(t)$ to obtain the superadiabatic basis \cite{adprob,berry,berry1,Fleischhauer1999}.
But for our purposes here, already the first order perturbative correction to the instantaneous adiabatic states of the system 
provides an accurate estimate of the reversible population leakage from the adiabatic ground state \cite{Benseny2021}.

\section{Application to the two-state Landau-Zener model}

We now compare the above perturbative approach for the calculation of the non-adiabatic transition probabilities 
with exact numerical results of the Landau-Zener model \cite{LZ1,LZ2}.  

%%%%%%%%%%%%%%%%%%%%%%%%%%%%%%%%%%%%%%%
\begin{figure}[t]
  \centerline{\includegraphics[width=0.8\linewidth]{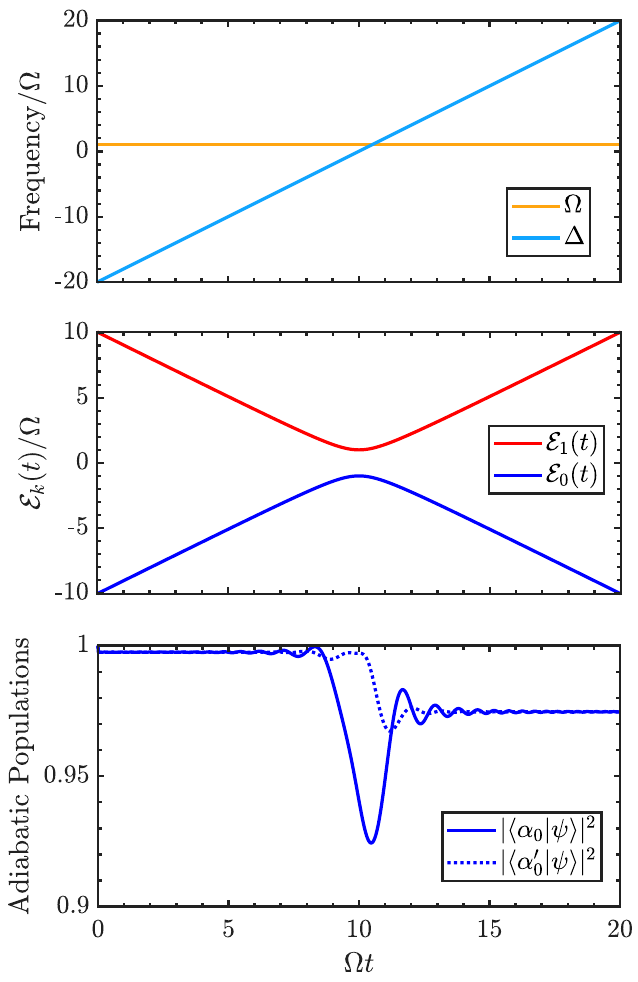}}
  \caption{Adiabatic passage in a two-level Landau-Zener model.
    The top panel shows the time-dependence of the Rabi frequency $\Omega$ and detuning $\Delta$ of the driving laser.
    The middle panel shows the time dependence of the energies $\mathcal{E}_{0,1}$ of the instantaneous eigenstates $\ket{\alpha_{0,1}(t)}$.
    The lower panel shows the populations of the adiabatic ground $\ket{\alpha_{0}}$ (blue solid line) of Eqs.~(\ref{eq:lz}) 
    and the dressed ground state $\ket{\alpha'_{0}}$ (blue dotted line) of Eq.~(\ref{eq:lzcorr}).}
  \label{fig:Lz}
\end{figure}
%%%%%%%%%%%%%%%%%%%%%%%%%%%%%%%%%%%%%

Consider a two-level atom with levels $\ket{g}$ and $\ket{e}$ interacting with a laser field 
with a constant Rabi frequency $\Omega$ and time-dependent detuning $\Delta(t)= b \, (t-T/2)$. 
The  Hamiltonian of the system ($\hbar = 1$) is 
\begin{equation}
	\mathcal{H}_{\mathrm{LZ}}(t) = 
        \left( 
        \begin{array}{cc}
	\Delta(t)/2 & -\Omega \\
	-\Omega & - \Delta(t)/2
	\end{array} 
        \right) .
\end{equation}
Solving the eigenvalue problem $\mathcal{H}_{\mathrm{LZ}} \ket{\alpha_{0,1}}=\mathcal{E}_{0,1} \ket{\alpha_{0,1}}$,
we obtain the instantaneous adiabatic eigenstates 
\begin{equation}
\ket{\alpha_{0,1}} = \frac{1}{\sqrt{M_{0,1}}} \left[ \left(\Delta/2 \pm  \sqrt{\Delta^2/4 + \Omega^2}  \right) \ket{g} 
- \Omega \ket{e} \right]
\label{eq:lz}
\end{equation}
with the corresponding energies $\mathcal{E}_{0,1} = \pm \sqrt{\Delta^2/4 + \Omega^2}$.

If we start with the atom in the ground state, $\ket{\psi(0)} = \ket{g} \simeq \ket{\alpha_{0}}$, and  
$|\Delta| \gg \Omega$, the dressed ground state is  
\begin{equation}  
\ket{\alpha_{0}'} \simeq \frac{1}{\sqrt{M}}  \left[ \ket{\alpha_{0}} + i \eta \ket{\alpha_1} \right] 
\label{eq:lzcorr}
\end{equation}
where $\eta = \frac{b \Omega}{8 (\Delta^2/4 + \Omega^2 )^{3/2}}$ has appreciable values $\sim b/(8\Omega^2)$ 
in the vicinity of the avoided crossing $|\Delta| \lesssim \Omega$ and is vanishingly small when $|\Delta| \gg \Omega$.

In Fig.~\ref{fig:Lz} we illustrate the dynamics of the two-level system for a time dependent $\Delta(t)$ and constant $\Omega$.  
The instantaneous eigenstates of $\mathcal{H}_{\mathrm{LZ}}(t)$ exhibit an avoided crossing of the energy levels $\mathcal{E}_{0,1}$ 
in the vicinity of $t=T/2$. Note that the population of state $\ket{\alpha_{0}}$ exhibits a dip near the avoided crossing 
where the energy level separation is smallest, but the population of the dressed state $\ket{\alpha_{0}'}$ remains large. 
Hence, some of the population leaked from the adiabatic ground state $\ket{\alpha_{0}}$ during the transfer
returns to that state at later times when the non-adiabatic perturbation vanishes and $\ket{\alpha_{0}'} = \ket{\alpha_{0}}$

%%%%%%%%%%%%%%%%%%%%%%%%%%%%%%%%%%%%
\begin{figure}[t]
  \centerline{\includegraphics[width=0.8\linewidth]{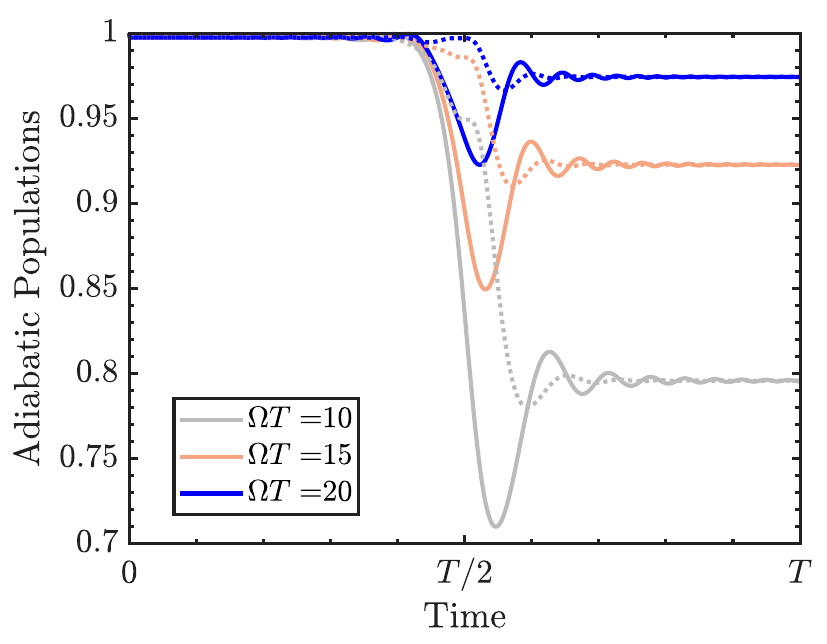}}
  \caption{Time-dependent population of the instantaneous adiabatic ground state $|\braket{\alpha_0|\psi}|^2$ (solid lines) and 
    the dressed ground state $|\braket{\alpha_0'|\psi}|^2$ (dotted lines of the same color), for different durations $T$ of the process.}
  \label{fig:2lsLZ}
\end{figure}
%%%%%%%%%%%%%%%%%%%%%%%%%%%%%%%%%%%%%

Finally, in Fig.~\ref{fig:2lsLZ} we show the populations of the adiabatic $\ket{\alpha_{0}}$ and dressed $\ket{\alpha'_{0}}$ ground states, 
for different durations $T$ of the adiabatic passage. This figure illustrates that the eigenstate dressing of Eq.~(\ref{eq:lzcorr}) describes 
the reversible population loss from the adiabatic state $\ket{\alpha_{0}}$ near the avoided level crossing, while shorter durations $T$, 
or larger rates of the frequency sweep $b$, lead to non-adiabatic and irreversible population loss from the state $\ket{\alpha_{0}}$ via 
the Landau-Zener transition to state $\ket{\alpha_{1}}$.


\begin{thebibliography}{100}
\bibitem{Matlab}
We use MATLAB standard command \texttt{eig()} to fully diagonalize the Hamiltonian (\ref{eq:ham}) for $N\leq 9$ atoms.
For larger systems, $N\leq 15$, we employ the Lanczos method to find the adiabatic ground and first excited states 
and their energy gap. 
For dynamical evolution, we construct the state propagation matrix using the eigenstates and eigenvalues of the system of 
$N\leq 9$ sites.
\end{thebibliography}
\end{document}